\documentclass[preprint,12pt]{elsarticle}

\usepackage{amssymb}
\usepackage{hyperref}
\usepackage{graphicx}
\usepackage{caption}
\usepackage{amsmath}
\usepackage{color}
\usepackage[symbol]{footmisc}

\journal{Results in Optics}
\begin{document}
\begin{frontmatter}



\title{Study of a bi-axial (KTP) crystal using Double Stokes Mueller Polarimetry}



\author{Chitra Shaji, Sruthil Lal S B, Alok Sharan \corref{cor1}}

\address{Department of Physics, Pondicherry University, Puducherry, India}
\ead{chitrashaji@gmail.com,getsruthil@gmail.com,*alok.phy@pondiuni.edu.in}

\begin{abstract}
We report the significance of the double Stokes Mueller polarimetry (DSMP) technique, to characterize a large size $(3\times3\times5 mm)$ KTP (Potassium titanyl phosphate) crystal. The crystal undergoes second harmonic generation with type II phase matching. The study of standard KTP crystal using the DSMP technique helps to validate the efficiency of this technique. We were able to extract the crystal's double Mueller matrix, relative contribution of the susceptibility tensor components, the phase difference between the susceptibility tensor components, etc. We could determine the crystal axes orientation using this optical technique, which was not possible through a single crystal X-Ray diffraction technique for such a large size crystal for which both optic axes and crystallographic axes are the same. Axes direction determined from polarization microscope measurements and Laue diffraction measurements on KTP crystal is compared with those obtained from DSMP measurements. 
\end{abstract}


\end{frontmatter}


\section{Introduction}
KTP crystal due to its large nonlinear coefficients, broad spectral bandwidth, high damage threshold, and large acceptance angle \cite{Shur2016,Nikogosyan2005,dezhong1985new}, is widely used for the second harmonic generation, sum and difference frequency generation, optical parametric amplification,  etc~\cite{Nikogosyan2005, mamrashev2018, Andrew1992, Driscoll1986,Vanherzeele:92,yaogang1986,belt1985ktp,driscoll1986efficient,anthon1988wavelength,bierlein892}. Recently, KTP crystal has also found application in terahertz wave generation \cite{mamrashev2018}. The varied applications of KTP crystal are possible due to the extensive study of its structural properties, crystal axes orientations, determination of 'd' coefficients, their relation to the physical properties of the crystal, etc \cite{Bierlein:89,bierlein1986electro,stolzenberger1988nonlinear}. We report, for the first time, the use of the double Stokes-Mueller polarimetry (DSMP) technique, to probe the second-order nonlinear optical properties of KTP crystal which is orthorhombic and biaxial in nature.\par
Double Stokes-Mueller polarimetry has recently emerged as a comprehensive technique to study nonlinear optical properties of materials especially to extract the second-order nonlinear optical coefficients of material \cite{Samim2015}. Linear Stokes-Mueller polarimetry (LSMP) has earlier been used with limited success to probe second-order nonlinear optical properties of KDP, quartz, ZnSe etc \cite{Mazumder2012, Lemaillet2007, Cisek2014}. The complex optical responses of biaxial crystals are also studied using LSMP \cite{Arteaga:19}. The theory of double Stokes-Mueller formalism based on particle nature of light was initially proposed by Y.Shi et.al \cite{Shi1994} and by Samim et.al, using the wave nature of light \cite{Samim2015}. The three-photon nonlinear optical process is also studied using nonlinear Stokes-Mueller formalism using particle as well as wave nature of light \cite{Shaji2017,Samim2016a}. This technique was effectively used to characterize the changes in ultrastructural features of the cancerous, tumorous tissues from that of normal tissues by way of comparing the different degree of linear polarization (DOLP) and finding the ratio of different susceptibility components \cite{Tokarz2015,Golaraei2016}. DSMP is used to study the dependence of depolarization on the thickness of the porcine samples \cite{ChukwuemekaOkoro2016, okoro2018second}; and by measuring the ratio of susceptibilities in retinal muscles of the eye in the fruit fly, a better understanding of the ultra-structural properties of the material is achieved. \cite{Samim2015,Kontenis2016}. The susceptibility ratios determined using DSMP give information about the structural properties of crystalline material in single starch granules as well as a lump of starch granules \cite{Samim2015,Cisek2017}. The entropy of the susceptibility tensors is determined for $z$-cut quartz crystal and collagen of rat tail tendon. The entropy of the susceptibility tensors sheds light on the organization of collagen in biological tissue \cite{Samim2017}. A wide-field second harmonic generation microscopy using DSMP technique is used to extract imaginary components of susceptibility tensors which in turn gives information regarding chiral susceptibilities of collagen fibers \cite{castano2019}. Monte Carlo polarization-sensitive SHG simulation model based on  DSMP is used to study nonlinear light interaction with tissues. This theoretical model prediction are compared with the DSMP measurements performed in rat tail collagen \cite{fung2018}. The R-ratio of myosin fibrils of Drosophila melanogaster larva is studied using incoming and outgoing circular polarization states in DSMP technique \cite{golaraei2020dual}.
Primarily the technique has been so far used to study diffusive systems and not much work has been done to explore it for its use in a single crystal. The advantage of this technique is in determining the degree of polarization thereby probing how do the different polarization states interact and propagate through the medium. This also has the potential to yield the information of orientation of crystallographic axes with respect to laboratory frame. This aspect is generally not dealt with as conventional techniques are used to determine so. As we experienced it is difficult to obtain this information for large-size crystals. We have extended the use of the DSMP technique to determine the orientation of the optic axes direction using the DSMP technique, besides obtaining the nonlinear optical coefficients. The KTP crystal is also analyzed using a polarization microscope as well as the Laue diffraction method. The results obtained from these techniques are compared with the results obtained through DSMP measurements.\par
The organization of the paper is as follows. We start with a brief description of the theory of DSMP in Section 2. Later, section 3 explains the experimental schematic of the DSMP technique. Section 4 presents the results obtained from DSMP, polarization microscope, and Laue diffraction measurements. Finally, the conclusions of the research work are explained in section 5.
\section{Theory}\label{theory}
 Double Stokes-Mueller polarimetry is a more comprehensive technique as compared to linear Stokes-Mueller polarimetry. In this technique, we take into account the nonlinear interaction of incident light with the material. The incident light propagating along $Y$ axis and polarization along $XZ$ plane in the laboratory frame is represented using $9\times1$ double Stokes vector. The second harmonic signal emergent from the material of study is represented using a $4\times1$ linear Stokes vector and the nonlinear optical properties of the material medium are represented using $4\times9$ double Mueller matrix \cite{Samim2015}. The general equation governing the relation between the incident light, emergent light, and optical properties of the material is shown in  Eq.\ref{stokesmuel},
 \begin{equation}
s(2\omega)=MS(\omega)
 \label{stokesmuel}
\end{equation}

where $S(\omega)$ is the $9\times1$ double Stokes vector of the incident light, $M$ is $4\times9$ double Mueller matrix of the material, and  $s(2\omega)$ is the $4\times1$ Stokes vector of emergent light. The double Stokes vector of the incident light is written in terms of its linear Stokes vector as shown below,
\begin{equation}
\label{doublestoke}
\begin{bmatrix}
S_1\\
S_2\\
S_3\\
S_4\\
S_5\\
S_6\\
S_7\\
S_8\\
S_9\\
\end{bmatrix}
=
\begin{bmatrix}
\frac{1}{\sqrt{6}}(3s_0^2-s_1^2)\\
\frac{1}{\sqrt{12}}(5s_1^2-3s_0^2)\\
-s_0s_1\\
\frac{1}{2}(s_2^2-s_3^2)\\
s_2(s_1+s_0)\\
-s_2(s_1-s_0)\\
-s_2s_3\\
s_3(s_1+s_0)\\
s_3(s_1-s_0)\\
\end{bmatrix}
\end{equation}
Where $S_{\alpha}(\alpha=1,9)$ are the nine components of the double Stokes vector written as superposition of the linear  Stokes vector $s_{\beta}(\beta=1,4)$ \cite{Samim2015}. The polarization states, as described by the Stokes vector, are chosen from the surface of Poincare sphere. It is used as a non-orthogonal basis vectors. These different polarization vectors excite different nonlinear susceptibility tensors and generate second harmonic signal. Measuring it in the orthogonal basis helps to determine the relation between the elements that combine to generate it. Since there could be superposition of light generated due to on or more elements, the resultant light would be undergo change in its ellipticity. This can be known by calculating various degree of polarization.  \par
The double Mueller matrix is calculated from the double Stokes vector of the incident light and the linear Stokes vector of the second harmonic generated light using the equation,
\begin{equation}
\label{doublemuel}
M=sS^{-1}
\end{equation}
Where, $S^{-1}$ is $9\times9$ matrix written by using the double Stokes vector of nine states of polarization incident on the material of study. These nine states of polarization are chosen from the surface of Poincare sphere and is written in terms of Poincare sphere coordinates as $S(0,0)$, $S(0,\frac{\pi}{2})$, $S(0,\frac{\pi}{4})$, $S(0,-\frac{\pi}{4})$, $S(\frac{\pi}{4},0)$, $S(-\frac{\pi}{4},0)$, $S(0,-\frac{\pi}{8})$, $S(\frac{\pi}{8},\frac{\pi}{2})$ and $S(\frac{\pi}{8}, \frac{\pi}{4})$. Similarly, $'s'$ in the equation \ref{doublemuel}, is the $4\times9$ matrix of the linear Stokes vector of emergent SHG light from the material. The linear Stokes vector of second harmonic generated light ($s$) gives information regarding the degree of polarization (DOP), degree of linear polarization (DOLP), degree of circular polarization (DOCP) (Eq \ref{DOP}).
\begin{equation}
 \label{DOP}
 DOP = \frac{\sqrt{s_{2}^2+s_{3}^{2}+s_{4}^{2}}}{s_{1}}\quad
 DOLP = \frac{\sqrt{s_{2}^{2}+s_{3}^{2}}}{s_{1}}\quad
 DOCP=\frac{\sqrt{s_{4}^{2}}}{s_{1}}
\end{equation}  
The double Mueller matrix elements obtained using  Eq.\ref{doublemuel}, is expressed in terms of the second order nonlinear optical susceptibility tensor components and its corresponding phase differences. The general expression showing relation between double Mueller matrix elements and nonlinear susceptibility tensor components is given below \cite{Samim2015}.
\begin{equation}
\label{muelsusceptib}
M_{\gamma N}=\frac{1}{2}Tr(\tau_\gamma\chi\lambda_N\chi^\dagger)
\end{equation}
where, $\tau_{\gamma}$ is set of Pauli matrices with $\gamma=1,4$. $\chi$ is the susceptibility tensors, $\lambda_{N}$ are the nine Gellmann matrices. N (N=1,9) is the subscript for the double Stokes vector as well as Gellmann matrices. Depending on the tensor components and its corresponding phase difference of the components involved, Mueller matrix elements can have "no-phase dependency", dependent only on the phase of "outgoing", or only on the phase of "incoming", or else phase dependency is on the combined "outgoing and incoming" components. Further, depending on the symmetry of the systems, some of the susceptibility tensor components are present and others vanish. Depending on that, some of the elements of the double Mueller matrix dominate or vanish. The nonlinear susceptibility tensor components deduced from the `no phase components' of the double Mueller matrix are given by following relations Eq.\ref{suscesix}.  Since we consider the light propagation along the $Y$ direction and polarization of light along $XZ$ plane, the susceptibility tensors having only $X$, $Z$ components would be present and others vanish.
\begin{equation}
\label{suscesix}
\begin{pmatrix}
|\chi^{(2)}_{ZXX}|^2\\
|\chi^{(2)}_{ZZZ}|^2\\
|\chi^{(2)}_{ZXZ}|^2\\
|\chi^{(2)}_{XXX}|^2\\
|\chi^{(2)}_{XZZ}|^2\\
|\chi^{(2)}_{XXZ}|^2\\
\end{pmatrix}
=
\begin{pmatrix}
\frac{1}{\sqrt{6}} & \frac{1}{2 \sqrt{3}} & \frac{1}{2} & \frac{1}{\sqrt{6}} & \frac{1}{2 \sqrt{3}} & \frac{1}{2} \\
 \frac{1}{\sqrt{6}} & \frac{1}{2 \sqrt{3}} & -\frac{1}{2} & \frac{1}{\sqrt{6}} & \frac{1}{2 \sqrt{3}} & -\frac{1}{2} \\
 \frac{1}{\sqrt{6}} & -\frac{1}{\sqrt{3}} & 0 & \frac{1}{\sqrt{6}} & -\frac{1}{\sqrt{3}} & 0 \\
 \frac{1}{\sqrt{6}} & \frac{1}{2 \sqrt{3}} & \frac{1}{2} & -\frac{1}{\sqrt{6}} & -\frac{1}{2 \sqrt{3}} & -\frac{1}{2} \\
 \frac{1}{\sqrt{6}} & \frac{1}{2 \sqrt{3}} & -\frac{1}{2} & -\frac{1}{\sqrt{6}} & -\frac{1}{2 \sqrt{3}} & \frac{1}{2} \\
 \frac{1}{\sqrt{6}} & -\frac{1}{\sqrt{3}} & 0 & -\frac{1}{\sqrt{6}} & \frac{1}{\sqrt{3}} & 0
\end{pmatrix}
\begin{pmatrix}
m_{1,1}\\
m_{1,2}\\
m_{1,3}\\
m_{2,1}\\
m_{2,2}\\
m_{2,3}
\end{pmatrix}
\end{equation}
\section{Experimental schematic}
\begin{figure*}[!htb]
\centering
\centerline{\includegraphics[width=14 cm, height=4 cm]{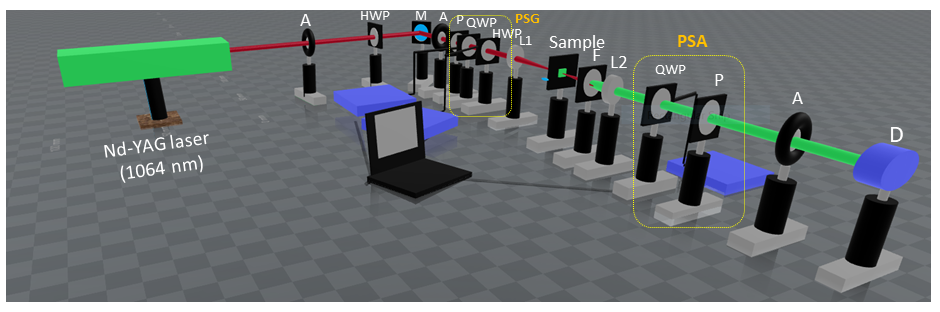}}
\caption{Experimental schematic of double Stokes-Mueller polarimeter used to study the second harmonic generation in a KTP crystal. Experimental setup consists of a Source (Nd-YAG laser, $1064nm$, $1W$), Aperture (A), Mirror (M), Polarization state generator (PSG), Lens (L1) of focal length = $20 cm$, Sample (Pottassium Titanyl Phosphate (KTP) crystal), Short pass filter (F) having cut-off wavelength = $600 nm$, Lens (L2) of focal length = $5 cm$ and Detector (D) (Silicon photo diode). PSG consists of Polarizer (P), Quarter-wave plate (QWP) and Half-wave plate (HWP) and PSA comprises of Quarter-wave plate (QWP) and  Polarizer (P)}
\label{dsmpktp}
\end{figure*}

 The schematic of DSMP performed on a KTP crystal to extract the nonlinear optical properties of crystal is shown in \autoref{dsmpktp}. Here we consider the propagation of light along the $Y$ axis and the polarization of light in $XZ$ plane in the laboratory frame coordinates. The experiment involves the generation of nine states of polarization incident on the material using PSG. The material of study, the KTP crystal ($3\times3\times5$ $mm$) is type II phase-matched nonlinear crystal which has an orthorhombic crystal structure and is biaxial in nature. The phase-matched crystal produces second harmonic generated light whose state of polarization is determined by measuring its linear Stokes vector. \par
The DSMP is implemented using cw Nd: YAG laser of wavelength $1064 nm$ (Diode-pumped laser kit, LASKIT-500) with $1W$ power as the source. The emission from the Nd-YAG laser initially passes through the aperture to keep the laser beam aligned with the rest of the optical layout (\autoref{dsmpktp}). The HWP following the aperture orients the state of polarization of the laser beam along the desired direction to get maximum power for the fundamental beam after the polarizer of PSG. The light entering into the PSG first passes through an aperture (A) which is used to align the reflected laser beam from the mirror (M) and also to block some of the reflected spots from the subsequent optical components. Then, light passes through the polarizer (10P309AR.16, Newport) oriented at $45^o$ with respect to its vertical axis. In the case of the other two elements of PSG, HWP (10RP52-3, Newport) and the QWP (10RP54-3, Newport), the fast axes are rotated in the clockwise direction from the vertical axis as we look into the source to produce nine states of polarization.  The waveplates of PSG are mounted on computer-controlled rotation mounts (SR50CC, SMC100CC, Newport), having repeatable accuracy of $0.03^{o}$. Rotation of waveplates of PSG generates nine states of polarization which are written in terms of Poincare sphere coordinates as $S(0,0)$, $S(0,\frac{\pi}{2})$, $S(0,\frac{\pi}{4})$, $S(0,-\frac{\pi}{4})$, $S(\frac{\pi}{4},0)$, $S(-\frac{\pi}{4},0)$, $S(0,-\frac{\pi}{8})$, $S(\frac{\pi}{8},\frac{\pi}{2})$ and $S(\frac{\pi}{8}, \frac{\pi}{4})$ \cite{Samim2015}. These nine states of polarization emerging from PSG pass through the uncoated lens (L1) which focuses the light onto the KTP crystal. The lens (L1) is mounted at a distance of $26 cm$  from the center of KTP crystal. The lens (L1) has a focal length of $20 cm$ at the visible region, but the same lens focuses the infrared light at a larger distance of about $26 cm$ at which the intensity of SHG shoots up. The second harmonic generated light is filtered from the emitted light from crystal by passing it through a short-pass filter (F) (cut-off wavelength = $600nm$) which allows only SHG light. The power of the SHG light is measured for each of the nine incident states of polarization by using a power meter (918D-UV-OD3) mounted next to the filter (This power meter is not fixed permanently in the experimental setup). The SHG light is collimated using a lens of focal length $5 cm$, to the short pass filter. We measure $4\times1$ linear Stokes vector of the second harmonic generated light ($532nm$) for all the nine incident states of polarization using the circular analyzer method \cite{Collett1984}.  The PSA consists of this circular analyzer constructed using a polarizer and quarter-wave plate.
The measurements performed using the polarizer of the PSA facing the source analyze all the linear polarization components of the SHG signal. The measurements carried out with the quarter-wave plate of PSA  facing the laser source give all the elliptical as well as the circular components of the SHG signal. The emergent light from PSA is incident on the silicon photodiode which is used to measure the intensity of SHG light.\par
    
The  KTP crystal on which the DSMP is performed is a phase-matched `external SHG crystal' used in the second harmonic generation of $1064nm$ in Laskit-500. The KTP crystal is aligned in such a way that the reflected spot of the crystal is aligned at a small angle along the XY plane throughout the experiment (\autoref{dsmpktp}). The measurements are carried out as the crystal is rotated from arbitrary $0^{\circ}$ to $180^{\circ}$ at an interval of $10^{\circ}$ in an anticlockwise direction about the direction of propagation of light ($Y$-axis) in a plane perpendicular to the direction of propagation ($XZ$ plane) as we look into the source. The linear Stokes vector of the SHG light is measured for nine incident states of polarization for each of the eighteen rotated positions of the KTP crystal. A total number of 36 measurements are carried out at each of the 18 different rotated positions of the crystal. After the measurement of linear Stokes vector of the emergent light from the crystal, the double Mueller matrix of the crystal is determined using \autoref{doublemuel}. The degree of polarization, the degree of linear polarization, and the degree of circular polarization of the SHG light are determined from the measured linear Stokes vector. The determination of the double Mueller matrix throws information on the susceptibility tensor components, the phase difference between the susceptibility tensor components, and the crystal axes orientations.  
\section{Results}
\subsection{Double Stokes-Mueller polarimetric measurements}
\begin{center}
\begin{figure*}[!htb]
\centerline{\includegraphics[width=\textwidth,height=6.5 cm]{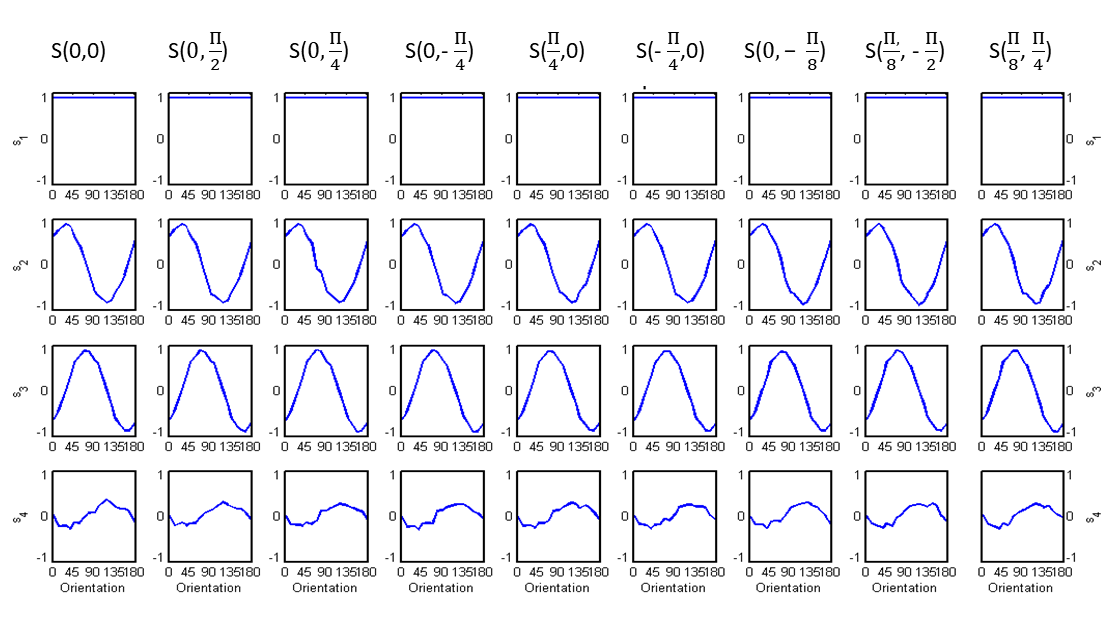}}
\caption{Variation of linear Stokes parameters (normalized in intensity) of the second harmonic generated light as a function of the angle of rotation of KTP crystal through $0^\circ$ to $180^\circ$ in a plane perpendicular to the propagation of light for the nine incident states of polarization. The state of polarization of the fundamental beam in terms of Poincare sphere coordinates is shown on the top of  column of the linear Stokes vector of corresponding SHG light.}
\label{lsktp}
\end{figure*}
\end{center}

 We have obtained the linear Stokes vector of the SHG light from the KTP crystal for the nine different incident states of polarization of fundamental light. It is measured as the crystal is rotated about the propagation direction of light ($Y$-axis) in a plane perpendicular to it ($XZ$ plane) (\autoref{lsktp}). The error in the measurement of the linear Stokes vector is estimated to be $5$ percent.
The percentage error in the elements of Stokes vector of SHG light is calculated using the formula given by Collet et.al., \cite{Collett1984}.
 \begin{equation}
\% error = S_0 (calculated)-S_0 (measured)/ S_0 (measured)
\end{equation}
 where $s_{0} (calculated) = \sqrt{s_1^2+s_2^2+s_3^2}$, where $s_1$, $s_{2}$, $s_{3}$ are experimentally determined Stokes parameters, $s_0 (measured)$ is also experimentally determined Stokes parameter. 
 Each column of \autoref{lsktp} shows the linear Stokes vector of second harmonic generated light for the nine states of polarization of the fundamental beam. We observe from \autoref{lsktp} that the pattern showing the variation of linear Stokes parameters remain the same irrespective of the different incident states of polarization, and the state of polarization of emitted SHG light represented by this linear Stokes vector spans from horizontal to vertical polarization with all other states of polarization in between them depending on the rotated position of the KTP crystal. The birefringence of the crystal attributes to the presence of `$s_{4}$' component of the linear Stokes vector (\autoref{lsktp}).
 A similar pattern showed by the elements of linear Stokes vector of SHG light for nine incident states of polarization is explained based on the biaxial nature of the KTP crystal. This biaxial crystal has two optics axes separated by 2V angle (2V=$37.4^o$ at 546.1nm) \cite{nikogosyan2006,Nesse2012}. This KTP crystal undergoes type II phase matching with `eoe' configuration where the fundamental beam is split in extraordinary (e) and ordinary (o) directions of the crystal and the SHG is polarized along the `e' direction \cite{Boyd2003}. The polarization direction of the SHG light obtained from the linear Stokes vector determines the direction of extraordinary refractive index ($n_{e}$) of the crystal as projected onto the XZ plane.
When the crystal is rotated about a plane perpendicular to the propagation direction, the direction of the optic axis and the direction of $n_{e}$ is also rotated. Due to this, the states of polarization of SHG light emerging from crystal span between horizontal and vertical. Whereas, for a fixed position of crystal, the state of polarization of SHG light remains the same for all the nine incident states of polarization. Though the state of polarization of the SHG light remains the same at a fixed position of crystal for all nine incident states of polarization, the power of SHG varies for all the incident states of polarization.\par
\begin{figure*}[ht]
 \centerline{\includegraphics[scale=0.5]{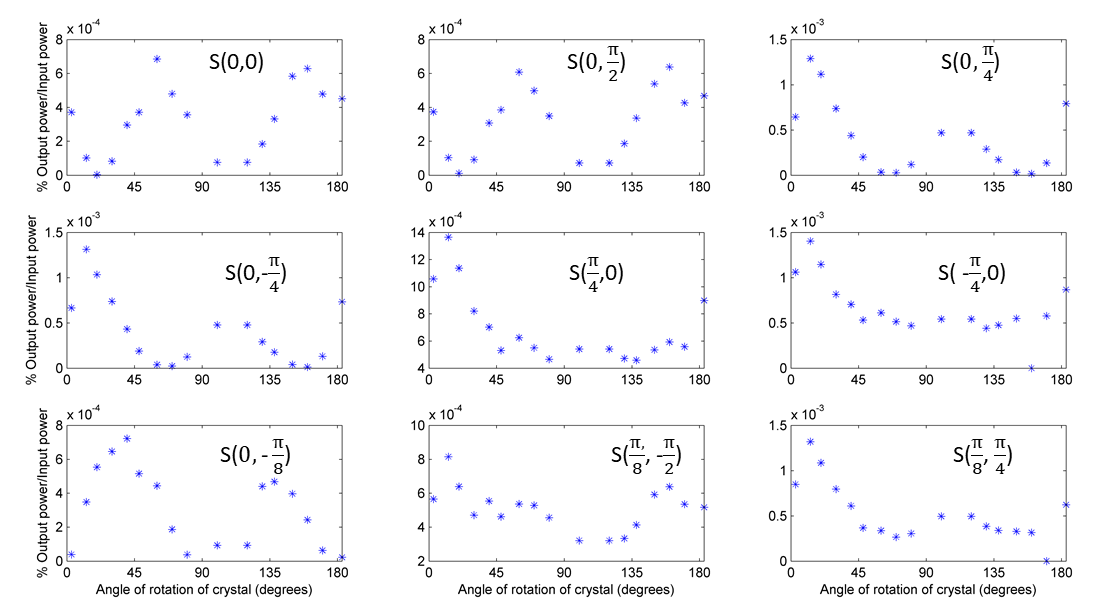}}
 \caption{Variation of the ratio of the power of the second harmonic generated (SHG) signal to the fundamental laser beam with angle rotation of KTP crystal through 0 to $180^\circ$ in a plane perpendicular to the propagation of light for nine incident states of polarization. The states of polarization of the fundamental beam are represented in terms of the Poincare coordinates inside each subplot.}
 \label{ktpeff}
 \end{figure*}

 The SHG efficiency measured as the ratio of the power of the SHG light and incident light is plotted as a function of the angle of rotation of crystal for each of the nine incident states of polarization (\autoref{ktpeff}).
From the  \autoref{ktpeff} it is clear that the SHG efficiency shows similar variation for the horizontal and the vertical states of polarization of the incident fundamental beam. Similarly, the SHG power variation for incident diagonal states of polarization as well as the circular states of polarization of fundamental beam show the same pattern (\autoref{ktpeff}). The variation of the SHG power for the other three incident states of polarization does not show similarity with each other (\autoref{ktpeff}). The power of the SHG light emitted from the crystal depends on the $d$ matrix elements of the KTP crystal \cite{Boyd2003}. When the crystal is illuminated with light of specific polarization, depending on the direction of orientation of the optic axis of the crystal, and also depending on the direction of polarization and propagation of incident light, only certain elements of the $d$ matrix are excited. Due to this fact, the intensity of SHG light which is dependent on these d matrix elements varies with the state of polarization of incident light for a fixed orientation of the KTP crystal.\par
\begin{figure*}[ht]
\centering
\includegraphics[scale=0.4]{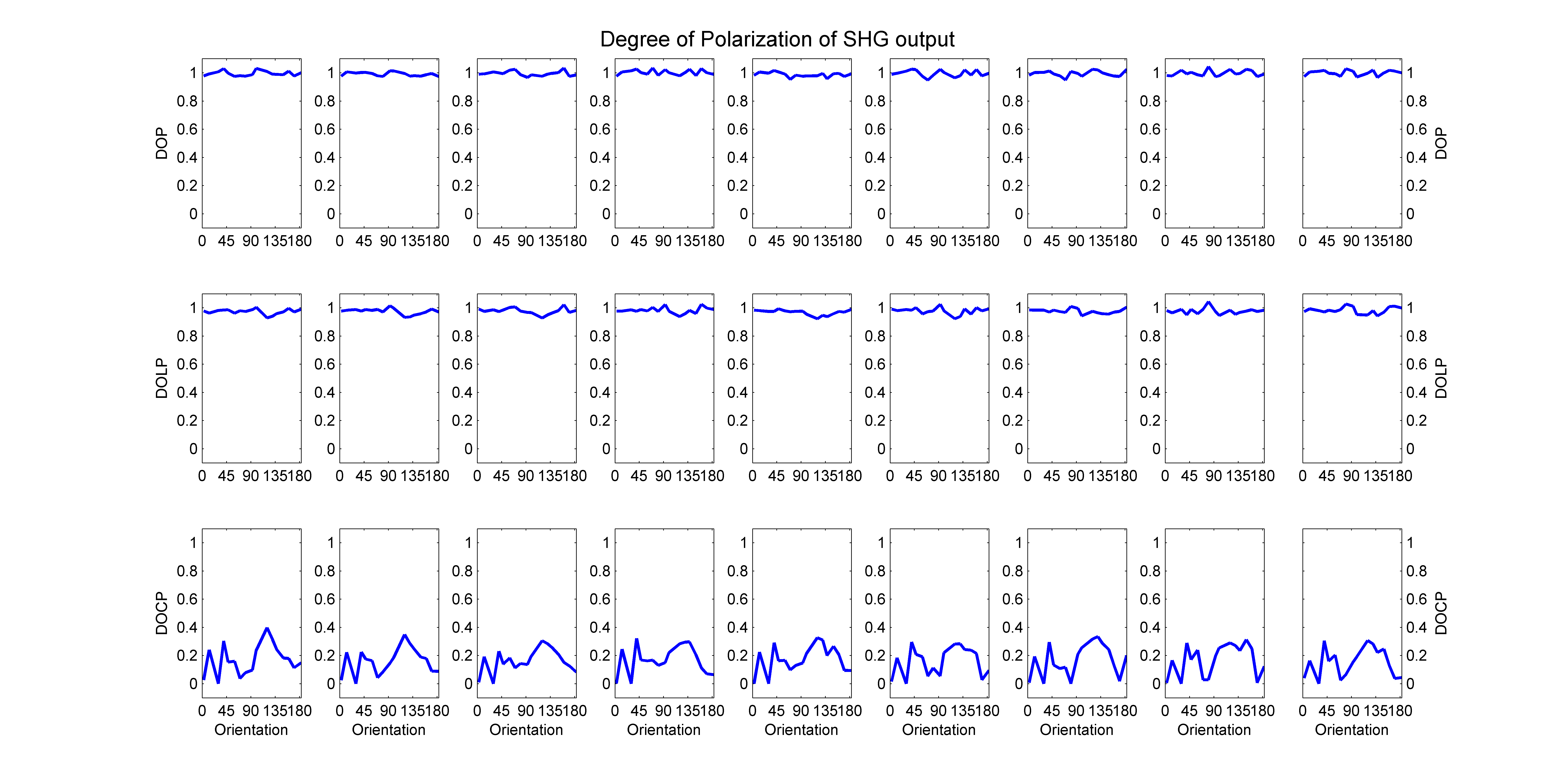}
\caption{Variation of degree of polarization (DOP), degree of linear polarization (DOLP), and degree of circular polarization (DOCP) as a function of the angle of rotation of KTP crystal in a plane perpendicular to the propagation of light for nine incident states of polarization. The state of polarization of the fundamental beam in terms of Poincare sphere coordinates is shown on the top of each column showing DOP, DOLP, and DOCP of the corresponding SHG light}
\label{dopktp}
\end{figure*}
 The degree of polarization (DOP) of the SHG light emitted from the crystal is calculated from its linear Stokes vector. The degree of linear polarization (DOLP) and degree of circular polarization (DOCP) are also calculated from the linear Stokes vector of SHG light (\autoref{dopktp}). The DOP, DOLP, and DOCP are plotted as a function of the angle of rotation of KTP crystal in a plane perpendicular to the propagation direction of the incident light (\autoref{dopktp}). DOP is a quantity that has a lump of information about the linear and the circular degree of polarization and it is validated from the experimental measurements (\autoref{dopktp}). This degree of polarization of light emerging from the material delivers information about the scattered light due to the medium. Since the system under study is a single crystal, the probability of depolarization of SHG light is less and it is observed from the variation of DOP in \autoref{dopktp}. The second and third rows in \autoref{dopktp} are the DOLP (Degree of linear polarization) and DOCP (Degree of circular polarization) of the SHG light respectively, which give us information about the linear polarization component and the circular polarization component of the SHG light. The birefringence of the crystal contributes to the DOCP. As we compare the second and third rows of \autoref{dopktp}, we can see that DOLP approaches one and DOCP has variations between 0 and 0.4. In addition to that, we have observed that the magnitude of DOCP shows an increase when  DOLP decreases.\par
 \begin{figure*}[!htb]
\centering
\centerline{\includegraphics[width=18cm, height=8cm]{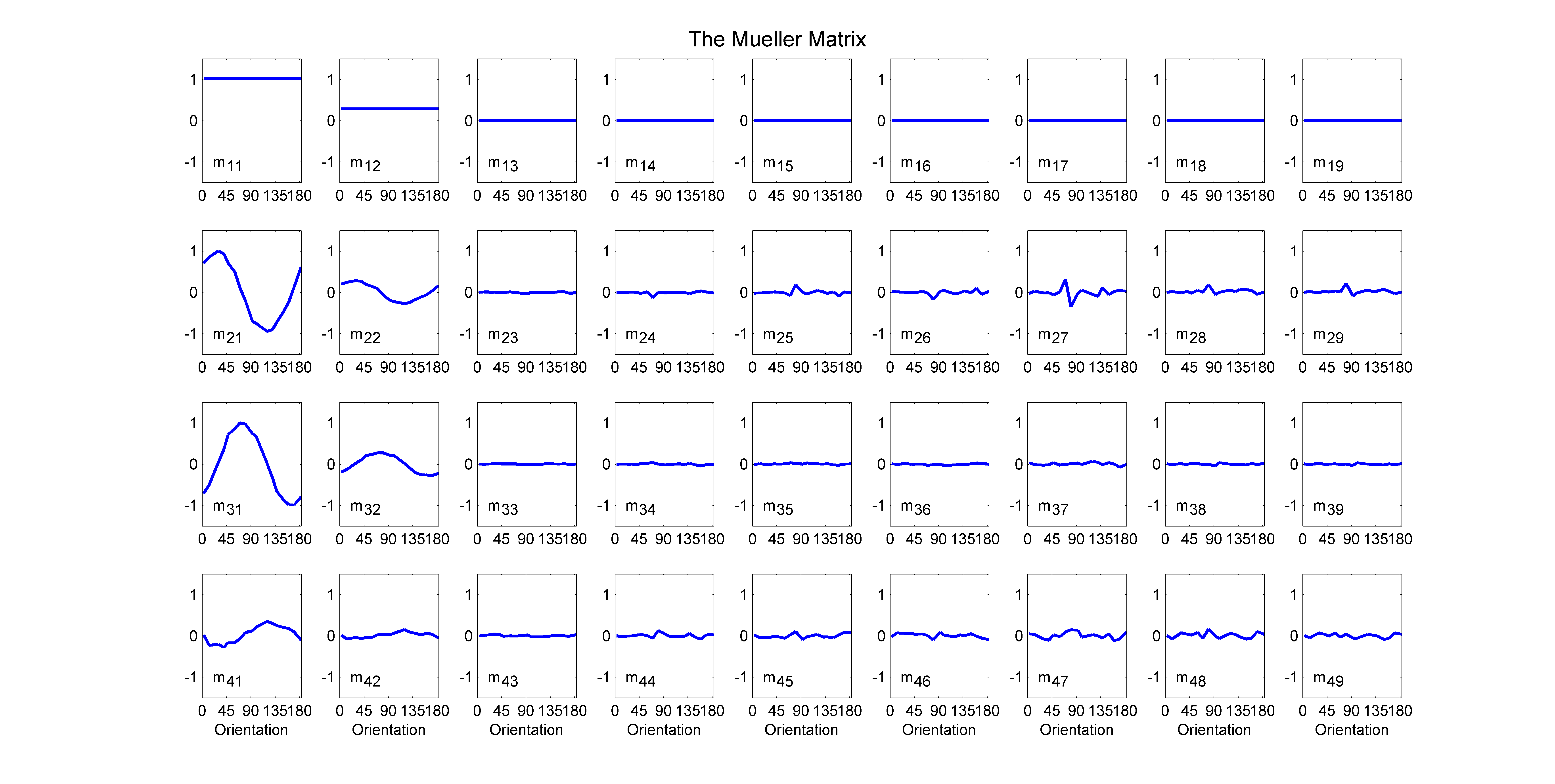}}
\caption{Variation of double Mueller matrix elements of KTP crystal with the angle of rotation of crystal from $0^\circ$ to $180^\circ$ in a plane perpendicular to the propagation of light. Each element ($m_{ij}$) of the double Mueller matrix is labeled as an inset of each of the plots.}
\label{ktpdm}
\end{figure*}
The double Mueller matrix of the KTP crystal is determined from the linear Stokes vector of the SHG light with the angle of rotation of the KTP crystal in a plane perpendicular to the propagation of light. The \autoref{ktpdm} shows the $4\times9$ double Mueller matrix with each element of it plotted as a function of the angle of rotation of the KTP crystal. Among the 36 elements of the double Mueller matrix, the first element $m_{11}$ is unity for all the angle of rotations of crystal, and the other most dominant elements in the double Mueller matrix are $m_{21}$ and $m_{31}$. The magnitudes of $m_{21}$ and $m_{31}$ vary from a maximum value of 1 to a minimum value of -1 as the crystal is rotated. The elements $m_{22}$, $m_{32}$, and $m_{41}$ of the double Mueller matrix show slight deviation from zero values at certain angles of rotation of the KTP crystal (\autoref{ktpdm}). From the \autoref{ktpdm}, it is visible that the elements $m_{11}$, $m_{21}$  and $ m_{22}$ are  non zero `No phase components' of the double Mueller matrix of this crystal. The element $m_{21}$ in \autoref{ktpdm} has a maximum value at $27^{0}$ of angle of rotation of the KTP crystal. At this angle of rotation of the crystal, the efficiency of SHG light emitted from the crystal approaches zero for incident horizontal and vertically polarized light, whereas the efficiency of SHG light for incident $+45^\circ$ and $-45^\circ$ polarized light is maximum. The minimum value of an element $m_{21}$ of double Mueller matrix is at an angle of  $117^\circ$, at this position also, the efficiency of the SHG light for incident horizontal and vertical states of polarization approaches to zero.  The $m_{22}$ of double Mueller matrix which is another `No phase component', has a maximum magnitude of 0.28 at $27^{\circ}$ and a minimum value of -0.27 at $117^{\circ}$. The non zero `O' components of the double Mueller matrix elements shown in \autoref{ktpdm} are $m_{31}$, $m_{41}$ and $m_{32}$. The magnitude of $m_{31}$ has a maximum value at $67^{\circ}$ and minimum value at $167^{\circ}$ of the angle of rotation of crystal. At an angle of $67^{\circ}$ of rotation of the KTP crystal, the efficiency of the SHG light is maximum for horizontal and vertical states of polarization and minimum for diagonal states of polarization of the fundamental beam incident on the crystal. The same trend is seen at $167^{\circ}$ rotated position of the crystal where the element $m_{31}$ has a minimum value. The variation of the element $m_{32}$ (non zero `O component') as a function of rotation of crystal shows the same trend as that of $m_{31}$ with a maximum value of 0.27 and a minimum value of -0.28 at the same angle of rotation of $m_{31}$. The element `$m_{41}$' which is another `O component' shows the minimum and maximum values at  $37^\circ$ and $117^\circ$  rotations of the KTP crystal respectively.\par 
\begin{figure}[htb]
\centering
\centerline{\includegraphics[scale=0.5]{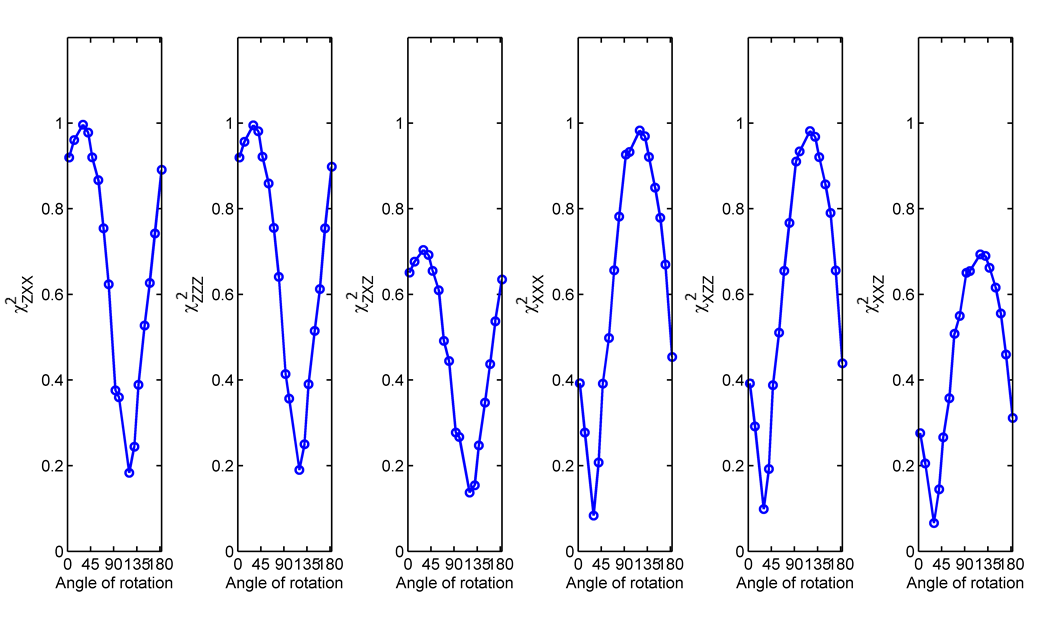}}
\vspace*{8pt}
\caption{Variation of relative contribution of second-order nonlinear susceptibility tensor components $\chi_{ZXX}^{(2)}$, $\chi_{ZZZ}^{(2)}$, $\chi_{ZXZ}^{(2)}$, $\chi_{XXX}^{(2)}$, $\chi_{XZZ}^{(2)}$ and $\chi_{XXZ}^{(2)}$ as a function of angle of rotation of the KTP crystal from 0 to $180^o$  in a plane perpendicular to the propagation of light}
\label{ktpsusc}
\end{figure}
\begin{figure}[!htb]
\centering
\centerline{\includegraphics[scale=0.4]{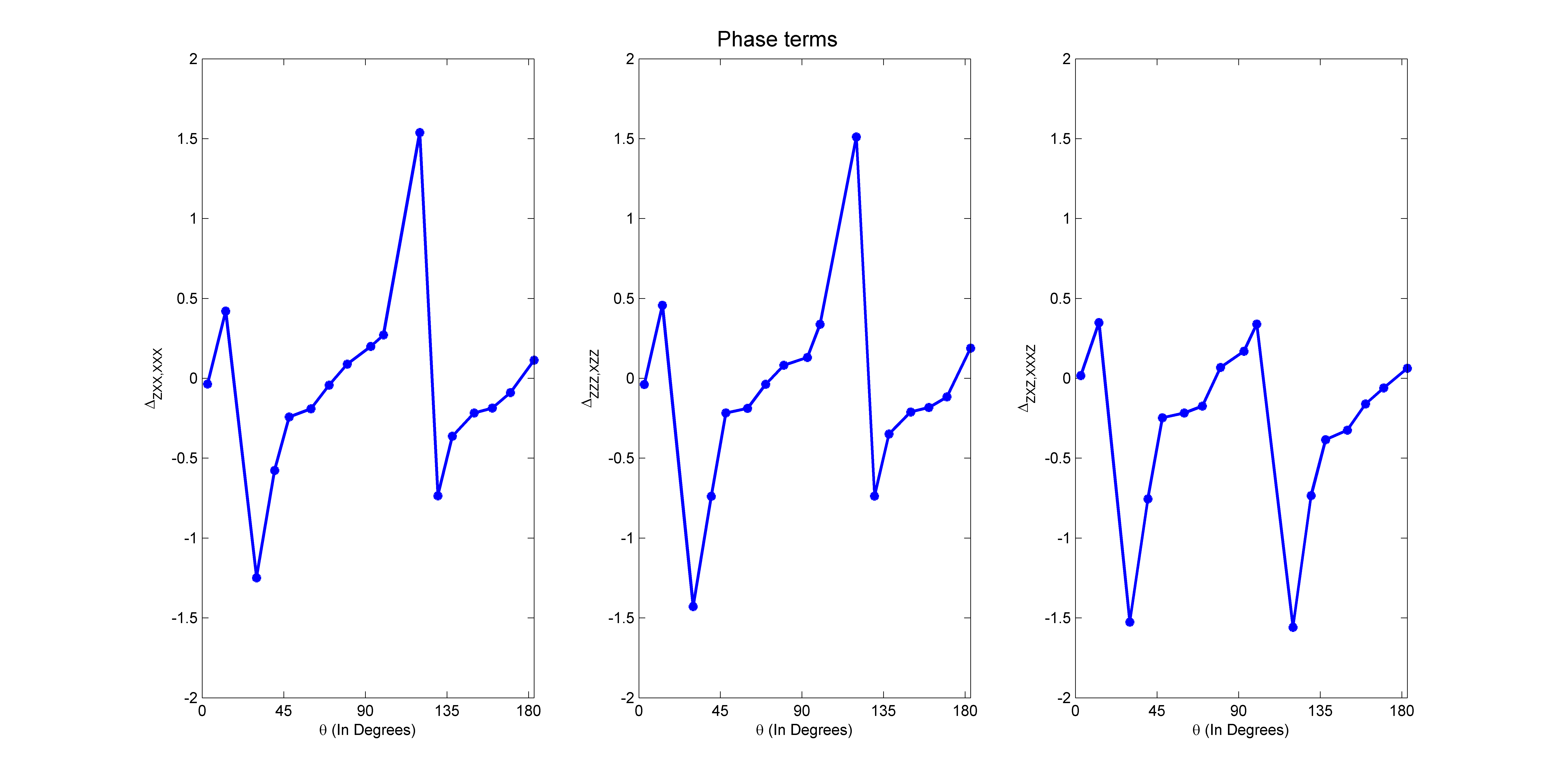}}
\vspace*{8pt}
\caption{Variation of phase differences between susceptibility tensor components of KTP crystal with the angle of rotation of crystal in a plane perpendicular to the propagation of light. Here, $\Delta_{ZXX,XXX}$ represents phase difference between $\chi_{ZXX}^{(2)}$ and $\chi_{XXX}^{(2)}$, $\Delta_{ZZZ,XZZ}$ represents phase difference between $\chi_{ZZZ}^{(2)}$ and $\chi_{XZZ}^{(2)}$, $\Delta_{ZZZ,XZZ}$ gives phase difference between $\chi_{ZZZ}^{(2)}$ and $\chi_{XZZ}^{(2)}$.}
\label{phasektp}
\end{figure}

  The relative contribution of nonlinear susceptibility tensor components of crystal are determined from the no phase components of double Mueller matrix elements \cite{Samim2015}. The determined susceptibility tensors of KTP crystal with the angle of rotation of it in a plane perpendicular to the direction of propagation are $\chi_{ZXX}^{(2)}$, $\chi_{ZZZ}^{(2)}$, $\chi_{ZXZ}^{(2)}$, $\chi_{XXX}^{(2)}$, $\chi_{XZZ}^{(2)}$ and $\chi_{XXZ}^{(2)}$ (\autoref{ktpsusc}). The \autoref{ktpsusc} shows the variation relative contribution of each of the susceptibility tensor components with the angle of rotation of the crystal. We observe from \autoref{ktpsusc} that the magnitude of $\chi_{ZXX}^{(2)}$, $\chi_{ZZZ}^{(2)}$, $\chi_{ZXZ}^{(2)}$ elements are maximum at an angle of rotation of $27^o$ of the crystal. The minimum magnitude of these nonlinear susceptibility tensor components is at an angle of rotation of $117^o$ of the KTP crystal. The minimum value for the other three susceptibility tensors, $\chi_{XXX}^{(2)}$, $ \chi_{XZZ}^{(2)}$ and $\chi_{XXZ}^{(2)}$ are at a rotated angle of $27^{o}$ of the crystal and the maximum value of these susceptibility tensors are at an angle of $117^o$. The phase difference between the susceptibility tensor components determined from the `O' components of the double Mueller matrix is plotted as a function of the angle of rotation of KTP crystal along the propagation direction of light in a plane perpendicular to it in \autoref{phasektp} \cite{Samim2015}. The \autoref{phasektp} depicts the relation between the phase difference of susceptibility components and the orientation of KTP crystal. \par
  The phase differences $\Delta_{ZXX,XXX}$ (phase difference between $\chi_{ZXX}^{(2)} and \chi_{XXX}^{(2)}$), $\Delta_{ZZZ,XZZ}$ (phase difference between $\chi_{ZZZ}^{(2)} and \chi_{XZZ}^{(2)}$) and $\Delta_{ZXZ,XXZ}$ have a dip at $27^o$, $\Delta_{ZXX,XXX}$ and $\Delta_{ZZZ,XZZ}$ (phase difference between $\chi_{ZZZ}^{(2)} and \chi_{XZZ}^{(2)}$) have a peek at $117^o$ whereas $\Delta_{ZXZ,XXZ}$ has a dip at $117^o$.\par
\subsection{Polarized light microscopy on KTP crystal}
\begin{figure}[!htb]
\centering
\centerline{\includegraphics[scale=0.55]{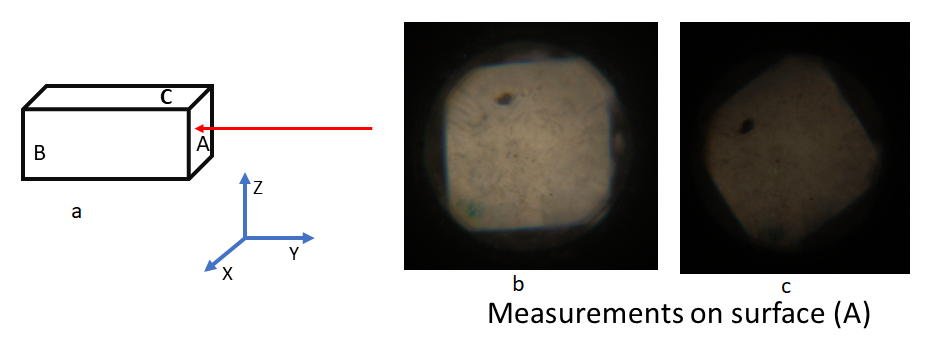}}
\caption{Polarized light microscopy images of the surface A of KTP crystal. Fig [a]: Surface A is illuminated with the light source where the propagation of light is along the $Y$-axis and the polarization of light is along the $XY$ plane in the laboratory frame coordinates. Fig [b] and Fig [c]: Polarized light microscopy yields bright (fig[b]) and dark images (fig[c]) of surface A of the crystal with the angle of rotation of $45^o$ about the propagation direction of light in a plane perpendicular to it.}
\label{polarmicroA}
\end{figure}

The KTP crystal is analyzed using the polarization microscope (Olympus BX-51) to determine the direction of optic axes. The polarized light microscopic studies are performed on three surfaces of the crystal (\autoref{polarmicroA}). The surface A of crystal is transparent, B and C are rough surfaces (fig[a] of (\autoref{polarmicroA})). The polarized light microscopy images of the transparent surface A with the rotation of crystal along the propagation direction of light ($Y$-axis) in a plane perpendicular to it, is shown in Fig [b] and Fig [c] of \autoref{polarmicroA}.

\begin{figure}[!htb]
\centering
\centerline{\includegraphics[scale=0.47]{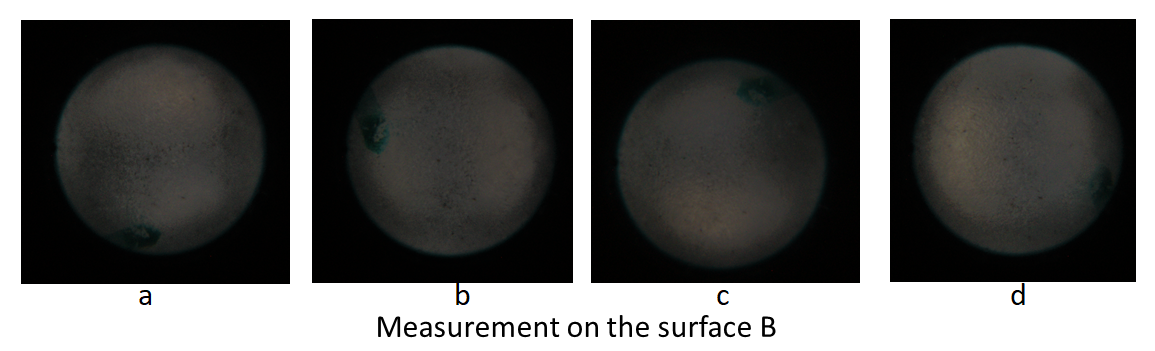}}
\caption{Polarized light microscopy images of the surface B of  crystal illuminated by the light source. The images show the presence of isogyres with the angle of rotation of the crystal by $90^\circ$ about the propagation of light in a plane perpendicular to it. The patterns in Fig [a], Fig [b], Fig [c] and Fig [d] depict the rotations of isogyres as the crystal is rotated by $90^\circ$.}
\label{polmicb}
\end{figure}

\begin{figure}[!htb]
\centering
\centerline{\includegraphics[scale=0.47]{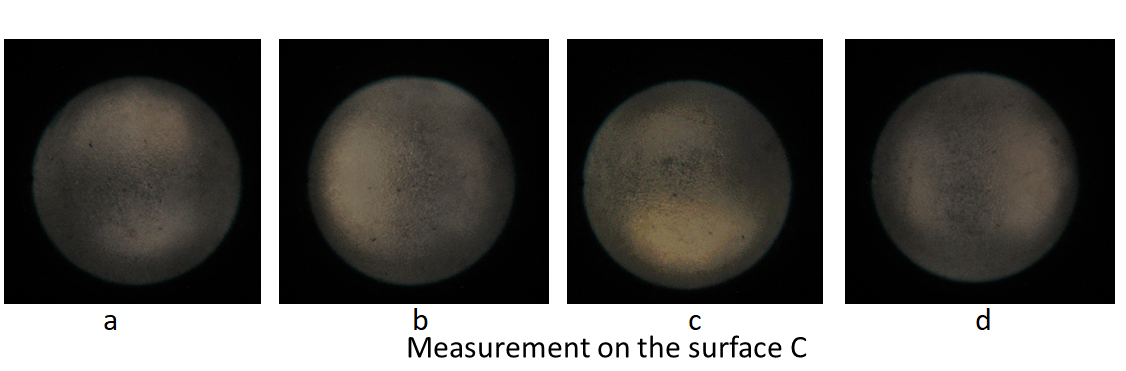}}
\caption{Polarization microscopy images for the illuminated  surface C of the crystal. The images show presence of isogyres with  the angle of rotation of crystal by $90^\circ$ about the propagation direction of light in a plane perpendicular to it. The patterns in Fig [a], Fig [b], Fig [c] and Fig [d] depict the rotations of isogyres as the crystal is rotated by $90^\circ$.}
\label{polmicc}
\end{figure}

 The transparent surface A of the crystal is illuminated by the light source (Halogen lamp). As the crystal is rotated for $45^{\circ}$ on the plane perpendicular to the plane of propagation of light, the intensity of light varies from bright to dark. A rotation of $90^{\circ}$ of the crystal brings back the bright field. The crystal behaves as a polarizing element which allows the incident polarized light for a particular orientation of it and diminishes when rotated by $45^o$. The polarized light microscopy images captured for surfaces B and  C are shown in \autoref{polmicb} and \autoref{polmicc} respectively.\par
 The presence of faint black lines that are rotated by the rotation of crystal is observed using a polarization microscope and is expected to be the isogyres. The bigger dimension of the crystal yields only faint images of the isogyres and the presence of these isogyres shows that optic axes point out from surfaces A and B \cite{Nesse2012}. Due to the larger dimension of the crystal, the optical path length travelled by light is larger as compared to the thin slices (typical $100 microns$) of crystal normally used for studies using polarization microscope. The polarized light microscopy images do not show too sharp a pattern as we expected, nevertheless, they help to know the orientation of optic axes of crystal approximately. 




 
\subsection{Laue diffraction in KTP crystal}
\begin{figure}[!htb]
\centerline{\includegraphics[scale=0.8]{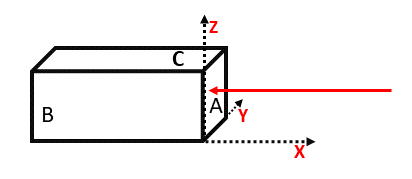}}
\caption{The laboratory frame coordinates used in performing Laue diffraction measurements. Light is incident on the surface A of crystal along the $X$-axis of lab frame. The [100] direction of the crystal is at an angle with respect to the long axis of the crystal.}
\label{laued}
\end{figure}

The Laue diffraction measurements are performed to determine the crystalline axes of the KTP crystal. The laboratory frame coordinates used for performing the Laue diffraction measurements are shown in \autoref{laued}.
The light beam is passed along the long axis of the crystal (propagation perpendicular to the surface A). The [100] direction is determined to be at an angle with respect to the physical crystal long axis. The goniometer used for the measurement of the angle of crystal is rotated by $17^\circ$ about $Z$ axis and $15^\circ$ about $Y$ axis. Still, an in-plane rotation of an angle of $45^\circ$ in the $XY$ plane is required for obtaining the [100] direction parallel to the $X$-axis in the laboratory coordinate. The measurement delivers the information that the crystal is not cut along any of the axes but rather is cut at a phase-matching angle for efficient SHG generation. 
\section{Conclusion}
The second harmonic generation process studied using the DSMP technique is an off-resonant process. The symmetry consideration for an off-resonant process allows the interchange of the indices of the susceptibility components. Hence, we expect the relative contribution of susceptibility components, $\chi_{ZXX}^{(2)}$ = $\chi_{XXZ}^{(2)}$ and $\chi_{ZXZ}^{(2)}$ = $\chi_{XZZ}^{(2)}$. From the  \autoref{ktpsusc} it is clear that the variation of susceptibility tensor components  $\chi_{ZXX}^{(2)}$ and  $\chi_{ZZZ}^{(2)}$, and $\chi_{XXX}^{(2)}$ and $\chi_{XXZ}^{(2)}$  are same. The \autoref{ktpsusc} also shows that the susceptibility tensor components $\chi_{ZXZ}^{(2)}$ and $\chi_{XXZ}^{(2)}$ have antisymmetric behavior. This discrepancy occurs since the susceptibility tensors of  KTP crystal determined from the double Mueller matrix measurements are done in the laboratory frame. The susceptibility tensors in the laboratory frame are a superposition of the susceptibility tensors of crystal in the molecular frame. The susceptibility tensor components $\chi_{ZXX}^{(2)}$ and $\chi_{ZZZ}^{(2)}$ of the crystal are the elements that have a maximum contribution to the generation of SHG light from a horizontally and vertically polarized fundamental beams respectively, these two elements contribute towards the generation of vertically polarized SHG signal. Since $\chi_{ZXX}^{(2)}$ and $\chi_{ZZZ}^{(2)}$ show identical behavior (\autoref{ktpsusc}), we understand that the crystal perceives the horizontal and vertical incident states of polarization to be same. This occurs when the unit cell of the crystal is tilted at an angle with respect to the physical crystal axes. The crystal identifies components of horizontal and vertical states of polarization to be equal when the crystal axis is rotated by $45^o$ with respect to the physical crystal long axis, which is exactly quantified by the Laue diffraction experiment performed on the KTP crystal. The other susceptibility tensor components also quantify this fact. The remaining two of the other susceptibility tensor components $\chi_{XXX}^{(2)}$ and $\chi_{XZZ}^{(2)}$ of the crystal show a similar pattern \autoref{ktpsusc}. These susceptibility components contribute to the generation of SHG light which is horizontally polarized. Since, the crystal views the horizontally and vertically polarized light as same, $\chi_{XXX}^{(2)}$ and $\chi_{XZZ}^{(2)}$ in the laboratory frame show the same variation. The remaining tensor components, $\chi_{ZXZ}^{(2)}$ and $\chi_{XXZ}^{(2)}$ are antisymmetric to each other \autoref{ktpsusc}. These tensor components are excited when the incident light is diagonally polarized. The estimation of susceptibility tensor components of the KTP crystal using DSMP helps to determine the orientation of the crystal axes, which makes DSMP an efficient technique to find out the crystal axes direction precisely.


\section{Acknowledgement}
We acknowledge Dr. Tamizhvel, Department of Condensed Matter Physics and Material Science, Tata institute of fundamental research, Bombay, for the Laue diffraction measurements of KTP crystal. We acknowledge Dr. Binoy Krishna Saha, Department of Chemistry, Pondicherry University for the discussions on single crystal X-Ray diffraction techniques. We also thank Dr. Rajneesh Bhutani, Professor, Department for Earth sciences, Pondicherry University for availing polarization microscope measurements. We thank Coordinator, Central Instrumentation Facility and Department of Physics, Pondicherry University  the use of laser and optical components. 

\bibliography{refproposal}

\begin{thebibliography}{10}

\bibitem{Shur2016}
V.~Y. Shur, E.~Pelegova, A.~Akhmatkhanov, and I.~Baturin, ``Periodically poled
  crystals of {{KTP}} family: a review,'' {\em Ferroelectrics}, vol.~496,
  no.~1, pp.~49--69, 2016.

\bibitem{Nikogosyan2005}
D.~N. Nikogosyan, {\em Nonlinear optical crystals: a complete survey}.
\newblock Springer Science \& Business Media, 2006.

\bibitem{dezhong1985new}
S.~Dezhong and H.~Chaoen, ``A new nonlinear optical crystal {{KTP}},'' {\em
  Progress in crystal growth and characterization}, vol.~11, no.~4,
  pp.~269--274, 1985.

\bibitem{mamrashev2018}
A.~Mamrashev, N.~Nikolaev, V.~Antsygin, Y.~Andreev, G.~Lanskii, and
  A.~Meshalkin, ``Optical properties of {{KTP}} crystals and their potential
  for terahertz generation,'' {\em Crystals}, vol.~8, no.~8, p.~310, 2018.

\bibitem{Andrew1992}
A.~J. Brown, M.~S. Bowers, K.~W. Kangas, and C.~H. Fisher, ``High-energy,
  high-efficiency second-harmonic generation of 1064-nm radiation in {{KTP}},''
  {\em Optics letters}, vol.~17, no.~2, pp.~109--111, 1992.

\bibitem{Driscoll1986}
T.~A. Driscoll, H.~J. Hoffman, R.~E. Stone, and P.~E. Perkins, ``Efficient
  second-harmonic generation in {{KTP}} crystals,'' {\em JOSA B}, vol.~3,
  no.~5, pp.~683--686, 1986.

\bibitem{Vanherzeele:92}
H.~Vanherzeele and J.~D. Bierlein, ``Magnitude of the nonlinear-optical
  coefficients of ktiopo4,'' {\em Opt. Lett.}, vol.~17, pp.~982--984, Jul 1992.

\bibitem{yaogang1986}
L.~Yaogang, X.~Bin, H.~Jianru, L.~Xiangyang, and J.~Minhua, ``Growth of
  ktiopo\_4 crystal for high efficiency shg devices and its main properties
  [j],'' {\em Chinese Journal of Lasers}, vol.~7, 1986.

\bibitem{belt1985ktp}
R.~Belt, G.~Gashurov, and Y.~Liu, ``{{KTP}} as a harmonic generator for {N}d:
  {YAG} lasers,'' {\em Laser focus (1983)}, vol.~21, no.~10, pp.~110--124,
  1985.

\bibitem{driscoll1986efficient}
T.~A. Driscoll, H.~J. Hoffman, R.~E. Stone, and P.~E. Perkins, ``Efficient
  second-harmonic generation in {{KTP}} crystals,'' {\em JOSA B}, vol.~3,
  no.~5, pp.~683--686, 1986.

\bibitem{anthon1988wavelength}
D.~W. Anthon and C.~Crowder, ``Wavelength dependent phase matching in {KTP},''
  {\em Applied optics}, vol.~27, no.~13, pp.~2650--2652, 1988.

\bibitem{bierlein892}
J.~D. Bierlein, A.~Ferretti, and M.~G. Roelofs, ``{KTiOPO4 ({KTP}): A New
  Material for Optical Waveguide Applications},'' in {\em Optoelectronic
  Materials, Devices, Packaging, and Interconnects II} (G.~M. McWright and
  H.~J. Wojtunik, eds.), vol.~0994, pp.~160 -- 167, International Society for
  Optics and Photonics, SPIE, 1989.

\bibitem{Bierlein:89}
J.~D. Bierlein and H.~Vanherzeele, ``Potassium titanyl phosphate: properties
  and new applications,'' {\em J. Opt. Soc. Am. B}, vol.~6, pp.~622--633, Apr
  1989.

\bibitem{bierlein1986electro}
J.~Bierlein and C.~Arweiler, ``Electro-optic and dielectric properties of
  ktiopo4,'' {\em Applied physics letters}, vol.~49, no.~15, pp.~917--919,
  1986.

\bibitem{stolzenberger1988nonlinear}
R.~A. Stolzenberger, ``Nonlinear optical properties of flux growth ktiopo 4,''
  {\em Applied optics}, vol.~27, no.~18, pp.~3883--3886, 1988.

\bibitem{Samim2015}
M.~Samim, S.~Krouglov, and V.~Barzda, ``Double {S}tokes {M}ueller polarimetry
  of second-harmonic generation in ordered molecular structures,'' {\em J. Opt.
  Soc. Am. B}, vol.~32, pp.~451--461, Mar 2015.

\bibitem{Mazumder2012}
N.~Mazumder, J.~Qiu, M.~R. Foreman, C.~M. Romero, C.-W. Hu, H.-R. Tsai,
  P.~T\"{o}r\"{o}k, and F.-J. Kao, ``Polarization-resolved second harmonic
  generation microscopy with a four-channel stokes-polarimeter,'' {\em Opt.
  Express}, vol.~20, pp.~14090--14099, Jun 2012.

\bibitem{Lemaillet2007}
P.~Lemaillet, S.~Rivet, F.~Pellen, B.~Le~Jeune, and J.~Cariou,
  ``Second-harmonic-generation-polarimeter calibration by means of a quartz
  plate,'' {\em Applied optics}, vol.~46, no.~21, pp.~4793--4803, 2007.

\bibitem{Cisek2014}
R.~Cisek, D.~Tokarz, N.~Hirmiz, A.~Saxena, A.~Shik, H.~E. Ruda, and V.~Barzda,
  ``Crystal lattice determination of {Z}n{S}e nanowires with
  polarization-dependent second harmonic generation microscopy,'' {\em
  Nanotechnology}, vol.~25, no.~50, p.~505703, 2014.

\bibitem{Arteaga:19}
O.~Arteaga and B.~Kahr, ``Mueller matrix polarimetry of bianisotropic materials
  (invited),'' {\em J. Opt. Soc. Am. B}, vol.~36, pp.~F72--F83, Aug 2019.

\bibitem{Shi1994}
Y.~Shi, W.~M. McClain, and R.~A. Harris, ``Generalized {S}tokes-{M}ueller
  formalism for two-photon absorption, frequency doubling, and hyper-raman
  scattering,'' {\em Phys. Rev. A}, vol.~49, pp.~1999--2015, Mar 1994.

\bibitem{Shaji2017}
C.~Shaji, R.~Ismail, S.~Satyanarayana, and A.~Sharan, ``Generalized {S}tokes
  vector for three photon process,'' {\em Journal of Nonlinear Optical Physics
  \& Materials}, vol.~26, no.~03, p.~1750040, 2017.

\bibitem{Samim2016a}
M.~Samim, S.~Krouglov, and V.~Barzda, ``Three-photon {S}tokes-{M}ueller
  polarimetry,'' {\em Physical Review A}, vol.~93, no.~3, p.~033839, 2016.

\bibitem{Tokarz2015}
D.~Tokarz, R.~Cisek, A.~Golaraei, S.~L. Asa, V.~Barzda, and B.~C. Wilson,
  ``Ultrastructural features of collagen in thyroid carcinoma tissue observed
  by polarization second harmonic generation microscopy,'' {\em Biomed. Opt.
  Express}, vol.~6, pp.~3475--3481, Sep 2015.

\bibitem{Golaraei2016}
A.~Golaraei, L.~Kontenis, R.~Cisek, D.~Tokarz, S.~J. Done, B.~C. Wilson, and
  V.~Barzda, ``Changes of collagen ultrastructure in breast cancer tissue
  determined by second-harmonic generation double {S}tokes-{M}ueller
  polarimetric microscopy,'' {\em Biomedical optics express}, vol.~7, no.~10,
  pp.~4054--4068, 2016.

\bibitem{ChukwuemekaOkoro2016}
K.~C.~T. Chukwuemeka~Okoro, ``Experimental demonstration of two-photon
  {M}ueller matrix second-harmonic generation microscopy,'' {\em Journal of
  Biomedical Optics}, vol.~21, pp.~21 -- 21 -- 6, 2016.

\bibitem{okoro2018second}
C.~Okoro, {\em Second-harmonic generation-based Mueller matrix polarization
  analysis of collagen-rich tissues}.
\newblock PhD thesis, University of Illinois at Urbana-Champaign, 2018.

\bibitem{Kontenis2016}
L.~Kontenis, M.~Samim, A.~Karunendiran, S.~Krouglov, B.~Stewart, and V.~Barzda,
  ``Second harmonic generation double {S}tokes {M}ueller polarimetric
  microscopy of myofilaments,'' {\em Biomed. Opt. Express}, vol.~7,
  pp.~559--569, Feb 2016.

\bibitem{Cisek2017}
R.~Cisek, D.~Tokarz, L.~Kontenis, V.~Barzda, and M.~Steup, ``Polarimetric
  second harmonic generation microscopy: An analytical tool for starch
  bioengineering,'' {\em Starch-St{\"a}rke}, 2017.

\bibitem{Samim2017}
M.~Samim, L.~Kontenis, S.~Krouglov, B.~C. Wilson, and V.~Barzda, ``Imaging of
  tissue disorder by the entropy of susceptibilities determined with second
  harmonic generation {S}tokes-{M}ueller polarimetric microscopy,'' in {\em
  Novel Techniques in Microscopy}, pp.~NTu3C--4, Optical Society of America,
  2017.

\bibitem{castano2019}
L.~J.~U. Castano, K.~Mirsanaye, A.~Golaraei, L.~Kontenis, R.~Cisek, H.~Zhao,
  and V.~Barzda, ``Wide-field polarimetric second harmonic generation
  microscopy of biological tissues,'' in {\em Nonlinear Optics}, pp.~NTu4A--5,
  Optical Society of America, 2019.

\bibitem{fung2018}
K.~B. Fung, M.~Samim, A.~Gribble, V.~Barzda, and I.~A. Vitkin, ``Monte carlo
  simulation of polarization-sensitive second-harmonic generation and
  propagation in biological tissue,'' {\em Journal of biophotonics}, vol.~11,
  no.~12, p.~e201800036, 2018.

\bibitem{golaraei2020dual}
A.~Golaraei, L.~Kontenis, A.~Karunendiran, B.~A. Stewart, and V.~Barzda,
  ``Dual-and single-shot susceptibility ratio measurements with circular
  polarizations in second-harmonic generation microscopy,'' {\em Journal of
  Biophotonics}, vol.~13, no.~4, p.~e201960167, 2020.

\bibitem{Collett1984}
E.~Collett, ``Measurement of the four {S}tokes polarization parameters with a
  single circular polarizer,'' {\em Optics communications}, vol.~52, no.~2,
  pp.~77--80, 1984.

\bibitem{nikogosyan2006}
D.~N. Nikogosyan, {\em Nonlinear optical crystals: a complete survey}.
\newblock Springer Science \& Business Media, 2006.

\bibitem{Nesse2012}
W.~D. Nesse, {\em Introduction to mineralogy}.
\newblock No.~549 NES, 2012.

\bibitem{Boyd2003}
R.~W. Boyd, {\em Nonlinear optics}.
\newblock Academic press, 2003.

\end{thebibliography}
\bibliographystyle{ieeetr}







\end{document}